# Polarization entanglement enabled by orthogonally stacked van der Waals NbOCl$_2$ crystals

Qiangbing Guo[1,2,#,*], Yun-Kun Wu[3,4,#], Di Zhang[5,7,#], Qiuhong Zhang[8], Guang-Can Guo[3,4], Andrea Alù[5,6], Xi-Feng Ren[3,4,*], Cheng-Wei Qiu[1,2,*]

[1]Department of Electrical and Computer Engineering, National University of Singapore, Singapore, Singapore.
[2]Centre for Advanced 2D Materials and Graphene Research Centre, National University of Singapore, Singapore, Singapore.
[3]CAS Key Laboratory of Quantum Information and CAS Synergetic Innovation Center of Quantum Information & Quantum Physics, University of Science and Technology of China, Hefei, China.
[4]Hefei National Laboratory, University of Science and Technology of China, Hefei, China.
[5]Photonics Initiative, Advanced Science Research Center, City University of New York, New York, USA.
[6]Physics Program, Graduate Center, City University of New York, New York, USA.
[7]The Key Laboratory of Weak-Light Nonlinear Photonics, Ministry of Education, School of Physics and TEDA Applied Physics Institute, Nankai University, Tianjin, China.
[8]School of Chemistry and Materials Science, Hangzhou Institute for Advanced Study, University of Chinese Academy of Sciences, Hangzhou, China.
[#]These authors contributed equally.
*qb_guo@nus.edu.sg
*renxf@ustc.edu.cn
*chengwei.qiu@nus.edu.sg

**Polarization entanglement holds significant importance for photonic quantum technologies. Recently emerging subwavelength nonlinear quantum light sources, e.g., GaP and LiNbO$_3$ thin films, benefiting from the relaxed phase-matching constraints and volume confinement, has shown intriguing properties, such as high-dimensional hyperentanglement and robust entanglement anti-degradation. Van der Waals (vdW) NbOCl$_2$ crystal, renowned for its superior optical nonlinearities, has emerged as one of ideal candidates for ultrathin quantum light sources [Nature 613, 53 (2023)]. However, polarization-entanglement is inaccessible in NbOCl$_2$ crystal due to its unfavorable nonlinear susceptibility tensor. Here, by leveraging the twist-stacking degree of freedom inherently in vdW systems, we showcase the preparation of tunable polarization entanglement and quantum Bell states. Our work not only provides a new and tunable polarization-entangled vdW photon-pair source, but also introduces a new knob in engineering the entanglement state of quantum light at the nanoscale.**

## Main

An entangled quantum light source serves as the foundation for various quantum technologies, including secure communications, quantum computing, quantum metrology and sensing, quantum imaging, as well as testing fundamental principles of quantum mechanics[1-7]. A second-order nonlinear ($\chi^{(2)}$) process, referred to as spontaneous parameter down conversion (SPDC), is the workhorse for generating such entangled quantum light states. Conventionally, this process involves the utilization of a bulky $\chi^{(2)}$ crystal with meticulously tailored crystal geometry to satisfy phase matching requirements[1,5,6]. However, commonly used $\chi^{(2)}$ crystals (such as BBO, KDP, KTP) are of low $\chi^{(2)}$ coefficients, necessitating large sizes for sufficient efficiency[2,3,4]. Recently, on-chip waveguides (e.g., periodically poled lithium niobate, PPLN) and microresonators (e.g. AlN microring, LN microdisk) have been developed and shown promise as bright SPDC sources[8-12], spurred by the advances in the fabrication and processing techniques of LN and AlN thin films. It is noted that either these conventional bulk crystals or on-chip waveguides/resonators adopts a light-matter interaction length (mm to cm levels) significantly longer than their coherent lengths, indicating a phase-matching condition is still indispensable for high efficiency or brightness, which, like bulk crystals, imposes similar restricts on the versatility and tunability of the generated quantum states[13].

Very recently, subwavelength thin SPDC sources have attracted broad attentions due to intriguing new features resulting from the relaxation of phase-matching constraints. These features include ultrabroad frequency and angular spectra, ultrashort correlation times and distances, as well as giant degrees of time-frequency and position-momentum entanglement[14-17]. This relaxation in subwavelength films enhances flexibility in engineering entangled quantum light states, essential for quantum optics and related technologies. For example, Santiago-Cruz et al. generated complex quantum states using a subwavelength semiconductor film with designed high-quality optical resonances[18], an approach not feasible with phase-matching engineered bulk crystals or on-chip waveguides/microresonators. These phase-matching relaxed subwavelength SPDC sources offer an unprecedented bridge between nanophotonics and quantum photonics, where the powerful abilities for light field manipulation (amplitude, polarization, optical mode, *etc.*) developed in nanophotonics will further enhance the tunability and versability of quantum light states that are challenging or impossible to achieve with large-size crystals or waveguides/miroresonators.

2D van der Waals (vdW) materials, known for their superior optical nonlinearities compared to conventional bulk $\chi^{(2)}$ crystals, in addition to their ease of vdW integration with various photonic structures[19-21]. These features make them an ideal platform for subwavelength SPDC sources. The high nonlinearity is particular critical for nonlinear devices with reduced interaction volume, while the ease vdW coupling with photonic structures greatly facilitate the integration of nanophotonics and quantum photonics to develop high-performance and multifunctional SPDC sources.

More importantly, as a unique enabling feature due to weak vdW bonding, twist engineering has been well established as a new knob for shaping and inducing exotic material properties, including superconductivity, correlated insulators, moiré excitons, topological phonon polariton transitions, optical nonlinearities, among many others[22-25]. For example, interlayer twist was recently shown to induce new second- and third-harmonic generation processes in 2D $WS_2$ via engineering of the atomic symmetry[26], and single photon emission from interlayer excitons in trapped moiré potential has been confirmed[27]. For SPDC processes in subwavelength film, the polarization configuration of the generated photon pairs is dictated by the nonlinear tensor structure[14]. VdW twist engineering is expected to potentially provide a novel approach to manipulate two-photon polarization states, which, however, has not yet been explored.

Very recently, $NbOCl_2$ has been identified as a vdW crystal with fairly high nonlinearity compared to other vdW materials and showcased the first ultrathin vdW SPDC source[28], offering a promising vdW SPDC platform at the nanoscale. However, as we have systematically revealed in this work, NbOCl$_2$ crystal intrinsically lacks polarization entanglement that is pivotal for practical qubit encoding[1,5,6,14,29,30]. To extend the function of this new vdW SPDC platform, we have further successfully constructed tunable polarization-entangled quantum states by resorting to NbOCl$_2$ vdW bi-layer orthogonal stacks. This presents a novel methodology for developing versatile, controllable, and compact vdW entangled quantum sources. Our work also implies vdW stack engineering can be a new knob for manipulating entangled quantum light states, facilitating an avenue to go further beyond the restrictions imposed by the intrinsic nonlinear tensor of natural materials and may provide higher freedom in engineering quantum states.

**Characterization of nonlinear tensor elements**

As NbOCl$_2$ crystal adopts a C2 symmetry[28], its second-order nonlinear susceptibility tensor can be represented as a 3×6 matrix

$$\chi^{(2)}_{ijk} = \begin{bmatrix} 0 & 0 & 0 & \chi^{(2)}_{abc} & 0 & \chi^{(2)}_{aab} \\ \chi^{(2)}_{baa} & \boldsymbol{\chi^{(2)}_{bbb}} & \boldsymbol{\chi^{(2)}_{bcc}} & 0 & \chi^{(2)}_{bac} & 0 \\ 0 & 0 & 0 & \boldsymbol{\chi^{(2)}_{cbc}} & 0 & \chi^{(2)}_{cab} \end{bmatrix}.$$

The second-order nonlinear processes are essentially generated by the interaction between the electric fields of light with the nonlinear polarization described by this $\chi^{(2)}$ tensor. Each tensor element corresponds to a specific combination of polarization directions and interactions within the material, dictating the efficiency and characteristics of the resulting nonlinear processes.

To gain deeper insights into the $\chi^{(2)}$ tensor elements, we performed polarization-resolved second harmonic generation (SHG) measurements in a transmission configuration (Figure 1a,b). Note that normal incidence onto the crystallographic basal plane (orthogonally defined by the *b*- and *c*-axes) is employed throughout this work (Figure 1b). In this pump geometry, the primary tensor elements governing the nonlinear processes are $\chi^{(2)}_{bbb}$, $\chi^{(2)}_{bcc}$ and $\chi^{(2)}_{cbc}$ (highlighted above in bold). Specifically, $\chi^{(2)}_{bbb}$ ($\chi^{(2)}_{bcc}$) defines the processes with pump polarized along the *b*-axis (*c*-axis) and emission polarized along the *b*-axis; $\chi^{(2)}_{cbc}$ defines the processes with pump polarized along the *b*- and *c*-axes and emission polarized along the *c*-axis.

As shown in Figure 1c, with an 808-nm pump, the SHG intensity defined by $\chi^{(2)}_{bbb}$ ($I_{bbb}$) is significantly larger than that defined by $\chi^{(2)}_{bcc}$ ($I_{bcc}$). The wavelength-dependent SHG intensity ratio of these two processes ($I_{bbb}/I_{bcc}$) is shown in Figure 1d. It is evident that the nonlinear process associated with $\chi^{(2)}_{bbb}$ is stronger than the others under normal pump conditions, gaining greater dominance at shorter wavelengths within the spectral range considered here.

To comply with the following SPDC experiments, we focus on the SHG process at 404 nm (with an 808-nm pump) hereinafter. More detailed azimuthal dependencies of SHG intensity under different polarization configurations were measured (Figure 1e; see Supplementary Figure S1 for additional data). These results were fitted using theoretical models to derive a tensor element ratio of $\chi^{(2)}_{bbb}/\chi^{(2)}_{bcc}$~4.65 (the calculation method is detailed in Supplementary Section S1). To account for the thickness dependence of the nonlinear processes, we modeled the nonlinearity intensity ratios ($I_{bcc}/I_{bbb}$ and $I_{cbc}/I_{bbb}$) based on this derived tensor element ratio (more details and analysis can be found in Supplementary Figure S2). The corresponding results are presented in Figure 1f, where the $\chi^{(2)}_{bbb}$ process exhibits significant dominance over the considered thickness range, and the weight gets higher with thickness. The modeled results are well-supported by the experimental data points presented in Figure 1f. Notably, the $\chi^{(2)}_{cbc}$-defined process is very low and significantly decreases with thickness (see detailed analysis in Supplementary Figure S2).

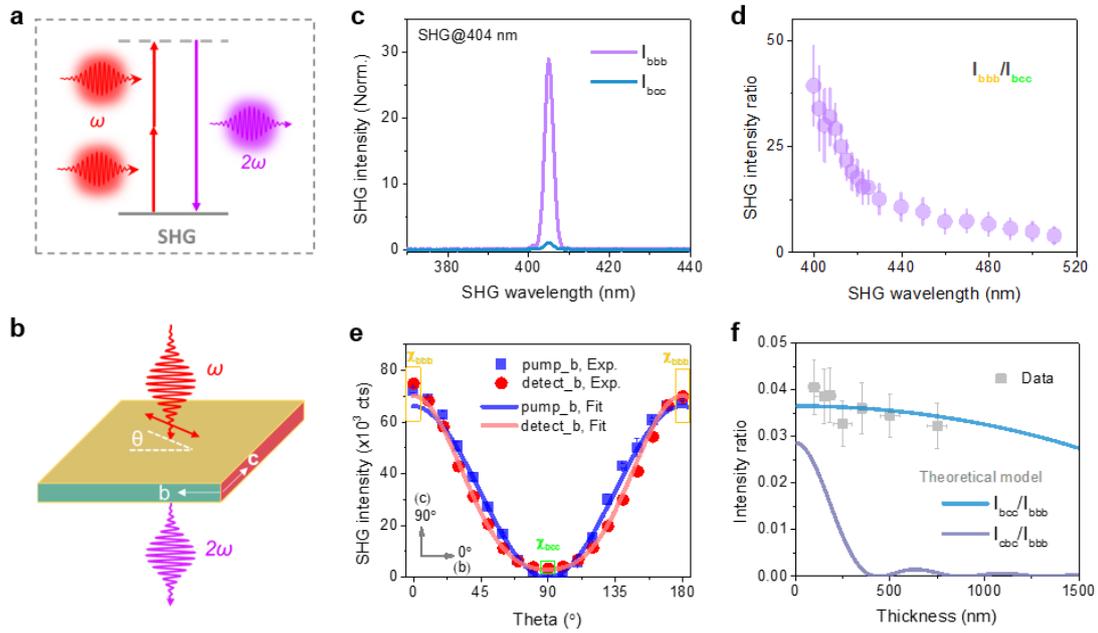

**Figure 1 | Characterization of nonlinear tensor elements through polarization-dependent SHG.** (a,b) Schematic illustration of the SHG process (a) and the experimental configuration (b). (c) Typical SHG spectra when pumped at 808 nm. (d) Wavelength dependent SHG intensity ratio between the $\chi^{(2)}_{bbb}$- and $\chi^{(2)}_{bcc}$-defined processes. (e) Azimuthally dependent SHG intensity by varying pump polarization. More details about the fitting can be found in the Supplementary Section S1. (f) Thickness dependence of nonlinear intensity ratios. The solid lines are modelling results based on the nonlinear tensor, while the dots are experimental results which are well consistent with the modelling. Error bars represent standard deviations from five measurements.

**SPDC process in single NbOCl$_2$ flakes**
SPDC is a $\chi^{(2)}$ process that can be seen as the time-reversal of SFG (Figure 2a), dictated by the same nonlinear tensor as in the SHG/SFG processes[31]. Therefore, the above-mentioned three tensor elements also define the main SPDC processes in transmission configuration as adopted in the following SPDC experiments (Figure 2b). As shown in Figure 2c, when pumped by a 404-nm continuous-wave (CW) laser in a Hanbury-Brown-Twiss interferometer, an obvious two-photon correlation peak with a $g^{(2)}(0)$ value over 260 could be observed with the pump and detection both polarized along the *b*-axis (i.e., $\chi^{(2)}_{bbb}$-defined process). This is a strong indicator for the generation of correlated photons, as uncorrelated photons result in accidental coincidences with no peak over delay time. Moreover, the inverse pump-power dependence of $g^{(2)}(0)$ value and the linear pump-power dependence of coincidence rates (Supplementary Figure S3), as well as the two-photon de Broglie wavelength measurement (Supplementary Figure S4), further unambiguously confirms the SPDC nature[13,15,18,29-32]. Notably, the photoluminescence signal in spectral region around 808 nm (i.e., the degenerate SPDC wavelength, see Supplementary Figure S5) is very weak, which underpins the measured high $g^{(2)}(0)$ value at a low pump power of ~1 mW.

By contrast, much weaker peak ($g^{(2)}(0)$~10) is with the $\chi^{(2)}_{bcc}$-defined process and no obvious peak with the $\chi^{(2)}_{cbc}$-defined one under the same pump power. Therefore, the $\chi^{(2)}_{bbb}$-defined process is also the dominating one in SPDC, with a small contribution from $\chi^{(2)}_{bcc}$-defined one and almost no contribution from $\chi^{(2)}_{cbc}$-defined one. Further polarization-dependent coincidence rates were measured (Figure 2d), which clearly indicates a strong polarization-dependence and is consistent with the above SHG analysis.

To gain a comprehensive picture of the polarization state of generated photon pairs in a single NbOCl$_2$ flake, we performed polarization tomographic measurement on a NbOCl$_2$ flake. As shown in Figure 2e, when pumped with a polarization of 45° (against the *b*-axis), the polarization state of the generated photon pairs exhibits an HH state (H denotes the polarization along the *b*-axis) with a fidelity of 0.95±0.02. Additional tomography results with a pump polarization of 0° could be found in Supplementary Figure S6, where the polarization of generated photon pairs has the same HH state. In other words, the polarization state of the generated photon pairs exhibits an HH state regardless of the pump polarization. Therefore, the generated photon pairs in single NbOCl$_2$ flakes are not polarization-entangled as limited by its crystal symmetry and thus nonlinear susceptibility tensor structure.

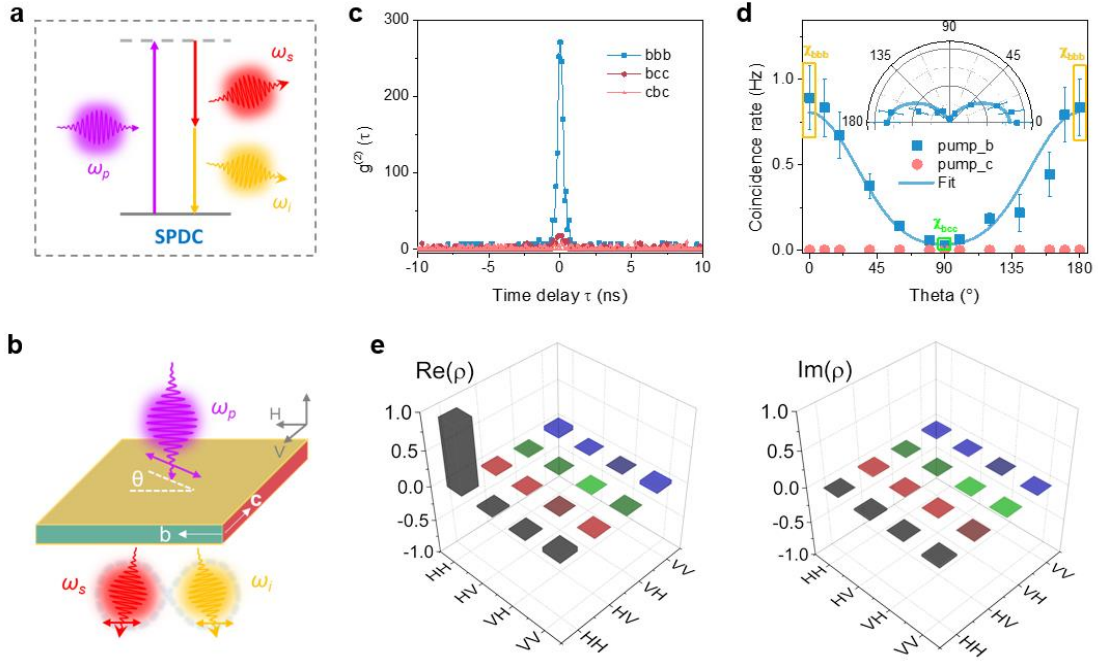

**Figure 2 | Polarization-dependent SPDC and quantum-state tomography in single NbOCl$_2$ flakes**. (a,b) Schematic illustration of SPDC process (a) and experimental configuration (b). (c) Normalized two-photon temporal correlation functions of three processes in a NbOCl$_2$ flake of ~200 nm thick under a pump power of ~1 mW with a 404-nm continuous-wave laser. Singal is filtered by a bandpass filter with a central wavelength of 810 nm and a full width at half maximal of 10 nm. (d) Polarization-dependent photon-pair coincidence rate. Inset is a presentation in polar coordinates. (e) Experimentally measured polarization density matrices for 45°-polarized pump. Error bars represent standard deviations from five measurements.

**Tunable polarization entanglement with NbOCl$_2$ bi-layer orthogonal stacks**

Given that polarization entanglement is pivotal for various quantum technologies[1,5,6,14, 29-32], the capability to generate polarization-entangled quantum states becomes a crucial step forward for such novel vdW SPDC sources. In principle, polarization entanglement can be obtained utilizing crystals with favorable nonlinear tensors[33,34]. For example, GaP and 3R-MoS$_2$ were recently demonstrated to generate polarization-entangled Bell states due to their cross-polarized nonlinear tensors[29,35]. While those materials have intrinsic access to polarization-entanglement guaranteed by specific crystal symmetry, the suitable nonlinear tensor types are essentially limited by available natural crystals, let alone the additional limitations imposed by their low nonlinearity (Supplementary Section S6). As compared and discussed in Supplementary Section S7, NbOCl$_2$ exhibits superior nonlinearity over GaP and 3R-MoS$_2$ thin films and conventional bulk crystals. Moreover, the SPDC efficiency in NbOCl$_2$ is also orders of magnitude higher than in the GaP thin-film and 3R-MoS$_2$ vdW SPDC sources (Supplementary Table S1). It would be ideal if polarization entanglement could be generated with this high-nonlinearity platform.

As discussed above, the generated photon pairs from a single flake remain a HH polarization state, regardless of pump polarization, as dictated by the fixed nonlinear tensor. It is natural to consider that if two pieces of crystals are stacked in an orthogonally twisted geometry (Figure 3a), a polarized pump with a non-zero angle ($\theta$) will down-convert in either crystal and these two possible SPDC processes will coherently superpose with one another as the spatial modes for the emitted photon pairs are indistinguishable for the stacked crystals[5,6]. As a result, this offers an opportunity to access controllable polarization states of the generated photon pairs through an elaborate combination of the geometry parameters ($d_1$, $d_2$ and θ, as defined in Figure 3a). In principle, the degree of polarization-entanglement can be widely tuned by manipulation of the weight balance between two paths (HH or VV, see Figure 3a). Specifically, maximally entangled Bell states can be prepared when the down-conversion occurs equally in either crystal.

We firstly developed a theoretical model to describe the effect of the geometry parameters of bi-layer stack on the polarization states of the generated photon pairs (see details in Method). The fidelity with two target Bell states (maximally entangled), i.e., $|\Phi^+\rangle = 1/\sqrt{2}(|HH\rangle+|VV\rangle)$ and $|\Phi^-\rangle = 1/\sqrt{2}(|HH\rangle - |VV\rangle)$, respectively, is presented in Figures 3b-e to reflect the two-photon polarization states. A fidelity value of 1 means a perfect Bell state ($|\Phi^+\rangle$ or $|\Phi^-\rangle$). Specifically, in the two-dimensional ($d_1$, $d_2$) parametric space, the mapping of available maximal fidelity with the two Bell states are shown in Figures 3b and 3c, respectively, where each point means the maximal fidelity value that can be obtained with the indicated thickness combination by scanning the pump polarization angle from 0° to 180°. Figures 3d and 3e show the fidelity value in the ($d_2$, θ) space, which represent the available fidelity by adjusting $d_2$ and θ when $d_1$ is fixed (tunability also exists in the ($d_1$, θ) space when $d_2$ is fixed). It is noted that the generated polarization states exhibit high tunability with the thickness combination (Figures 3b and 3c) and pump polarization angle (Figures 3d and 3e), which is also reflected in the high tunability of the degree of polarization entanglement as evidenced by the mapping of concurrence with the geometry parameters in Supplementary Section S8. These tunable entangled states are of fundamental interests for experimental tests of quantum mechanics[36-38], as well as of technological relevance for enhanced quantum sensing, metrology, and simulation due to being more robust to certain types of noise and decoherence[39-42].

More importantly, the pump polarization angle θ could serve as a post-fabrication knob for tuning the polarization entanglement states, i.e., when stacks are fabricated ($d_1$ and $d_2$ are fixed). Take the stack case ($d_1$=80 nm, $d_2$=260 nm) as an example, the fidelities with both Bell states can be widely tuned between zero to unity by simply controlling the pump polarization (Figure 3f), facilitating a facile access to tunable polarization-entangled states. In addition, switching between the two Bell states could be achieved by adjusting the pump polarization angle.

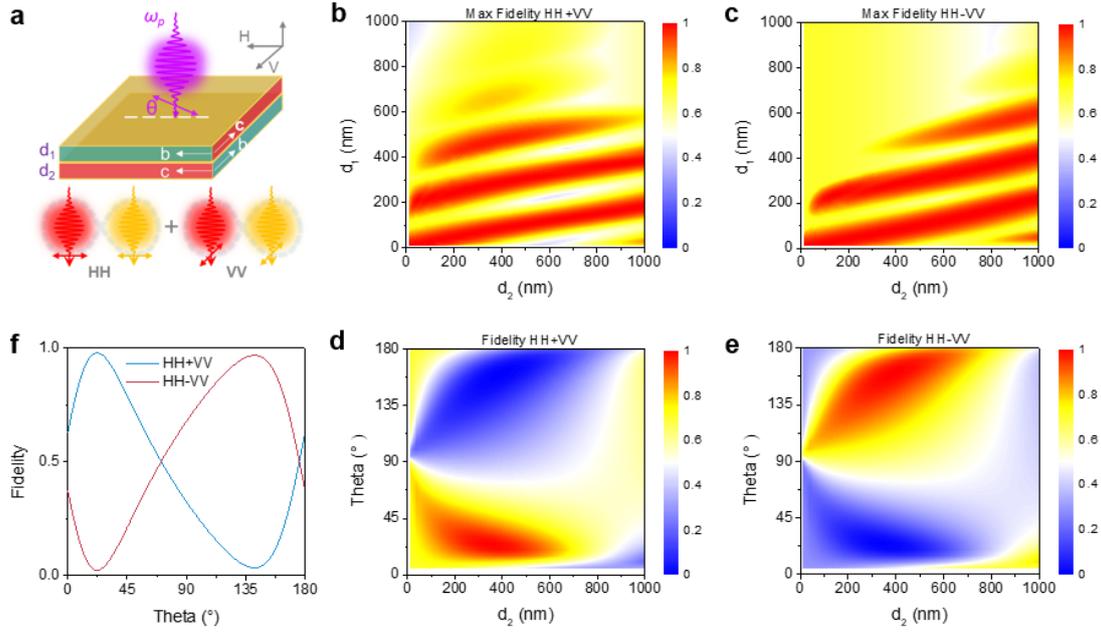

**Figure 3 | Tunable polarization entanglement with NbOCl$_2$ bi-layer orthogonal stacks**. (a) Schematic illustration of the geometry of the orthogonally twisted vdW bi-layer stack. $d_1$ ($d_2$) is the thickness of the first (second) crystal layer encountered by the incident pump. $\theta$ is the pump polarization angle defined against the crystallographic b-axis of the first crystal layer. (b,c) Calculated maximal fidelity in ($d_1$, $d_2$) parametric space, with two Bell states, $|\Phi^+\rangle = 1/\sqrt{2}(|HH\rangle+|VV\rangle)$ (HH+VV) and $|\Phi^-\rangle = 1/\sqrt{2}(|HH\rangle - |VV\rangle)$ (HH-VV), respectively. Each point in the parametric spaces represents the maximal fidelity that can be obtained by scanning the pump polarization angle $\theta$. (d,e) Calculated fidelity with the two Bell states in ($d_2$, $\theta$) parametric space with $d_1$ fixed at 80 nm. (f) Dependence of fidelity on pump polarization angle for the bi-layer stack case with $d_1$=80 nm and $d_2$=260 nm.

**Experimental demonstration and construction of Bell states**

As proof-of-concept demonstrations, we firstly fabricated a bi-layer stack ($d_1$~80 nm, $d_2$~260 nm) by picking up two exfoliated flakes of target thickness and assembling them into an orthogonal stack with all-dry transfer technique (Figure 4a, see Method for details). To gain an insight into the pump-polarization-dependent polarization state of the generated photon pairs, we measured the coincident rates along the HH and VV paths (Figure 3a), respectively, by scanning the pump polarization angle from 0° to 180°. The results are shown in Figure 4b, which indicates the variation of generated photon pairs with pump polarization in the two polarization paths and balances can be reached at ~25° and ~140°, respectively. This is well consistent with the modeled results shown in Figure 3f, where maximally entangled Bell states could be accessed near these balanced pump-polarization angles.

To reconstruct the polarization state of the generated photon pairs, polarization tomographic measurements were conducted (Figure 4c and see Method for more details). The density matrix $\rho$ of the polarization state can be derived by projecting into 16 different basis states. As shown in Figures 4d, the reconstructed density matrix with a pump polarization angle of 25° is presented and its fidelity with the $|\Phi^+\rangle = 1/\sqrt{2}(|HH\rangle+|VV\rangle)$ Bell state (Figure 4e) is derived to be 0.91±0.02 (see Supplementary Section S9 for more details). Besides, the reconstructed density matrix with a pump polarization angle of 140° is presented in Figure 4f and its fidelity with the $|\Phi^-\rangle = 1/\sqrt{2}(|HH\rangle - |VV\rangle)$ Bell state (Figure 4g) is derived to be 0.92±0.03. These results align well with the modeled ones (Figure 3f).

Furthermore, additional stacks with different thickness combinations were fabricated and their polarization entanglement states were measured, of which the reconstructed polarization density matrices are presented in Supplementary Section S10. It is noted that polarization-entangled Bell states can be obtained by varying thickness combinations and pump polarization, as well described by the developed theoretical model.

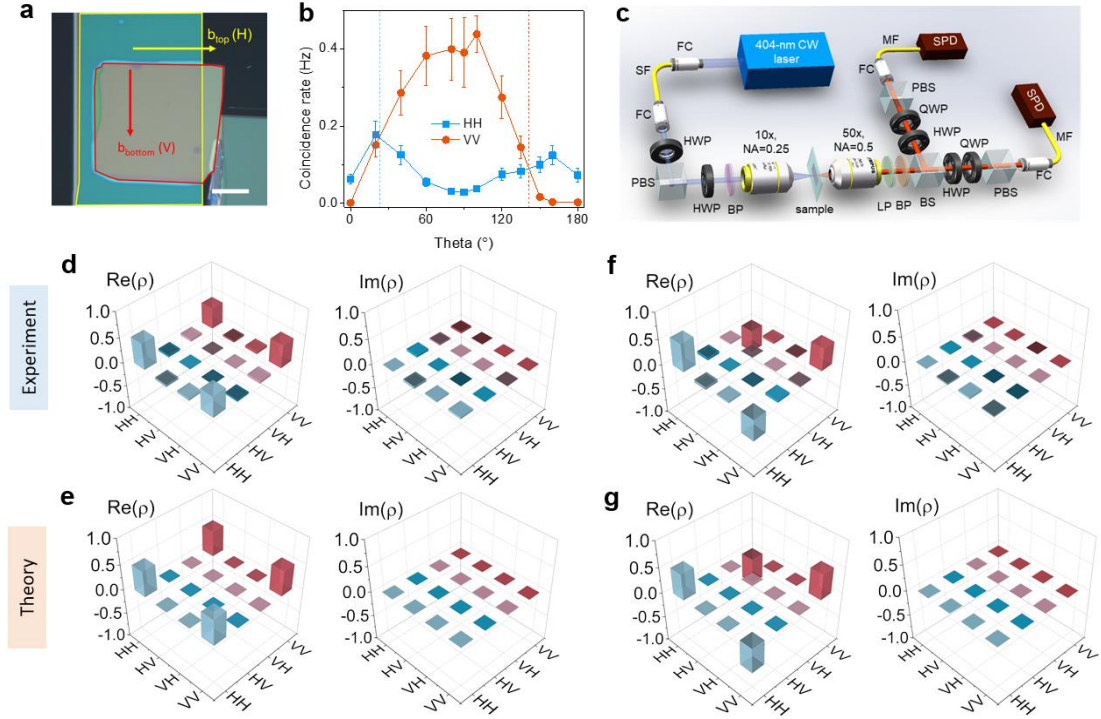

**Figure 4 | Experimental demonstration and Bell state construction with vdW bi-layer orthogonal stacks.** (a) Optical image of a fabricated orthogonal vdW stack. $d_1$~80 nm, $d_2$~260 nm. Scar bar, 25 μm. (b) Coincidence rate of two paths (HH and VV) as a function of pump polarization angle. Error bars represent standard deviations from three measurements. (c) Illustration of the optical setup for polarization quantum state tomography. FC, fiber coupler; SF, single-mode fiber; PBS, polarizing beam splitter; HWP, half-wave plate; QWP, quarter-wave plate; BP, band-pass; LP, long pass; BS, polarization-independent beamsplitter; SF, single-mode fiber; MF, multimode fiber; SPD, single-photon detector. (d,f) Experimentally measured polarization density matrices ρ for pump polarization angle θ of 25° (d) and 140° (f), respectively. The excitation power incident on samples is ~4 mW. (e,g) Theoretical density matrices for two Bell states, i.e., $|\Phi^+\rangle = 1/\sqrt{2}(|HH\rangle+|VV\rangle)$ (e) and $|\Phi^-\rangle = 1/\sqrt{2}(|HH\rangle - |VV\rangle)$ (g), respectively.

**Conclusion and perspective**

In conclusion, we have demonstrated the generation of tunable polarization-entangled quantum light states based on bi-layer orthogonal stacks of NbOCl$_2$ crystal, a recently discovered high-nonlinearity $\chi^{(2)}$ vdW crystal without inherent polarization entanglement. In addition, we have successfully constructed polarization-entangled Bell states with fidelities >0.9. Our work not only provides a novel subwavelength polarization-entangled photon-pair source with high-tunability, but also introduces a methodology for constructing polarization-entangled quantum light states at the nanoscale from crystals inherently lacking polarization entanglement which has the potential to significantly expand the toolbox for engineering the two-photon polarization entanglement states.

While we present a proof-of-concept demonstration, the fidelity of the polarization-entangled Bell states can be further enhanced by more precise control of sample thickness, stacking angle, and pump polarization angle. Exciting opportunities are also anticipated to emerge in creating and manipulating quantum light states through interfacing this vdW platform with cavity photonics and metaoptics[43-45], in light of the facile vdW integration with various photonic structures and twist-stacking degree of freedom demonstrated in this work.

As a novel platform for engineering photon-pair states at the nanoscale, greater tunability with this twistable vdW platform is expected. For example, while we have demonstrated tunability in bi-layer orthogonal stacks, additional tunability could be accessible by employing arbitrary twist angles and by resorting to tri-layer stacks, as shown and discussed in Supplementary Section 11. In addition, arbitrary qutrit states could also be generated with tri-layer stacked structures (see Supplementary Figure S13 for details), underpinning an avenue towards the preparation of complex quantum states at the nanoscale that are challenging for a natural $\chi^{(2)}$ crystal with a fixed nonlinear tensor[30,46].

## Methods

**Crystal synthesis and characterizations**. The single crystals were synthesized by chemical vapor transport as detailed in our recent paper[26]. The crystalline phase was confirmed with X-ray diffraction (Rigaku D/MAX 2550/PC with Cu Kα X-rays). Atomic force microscopy (Bruker Dimension Fast Scan in tapping mode) was used to identify the thickness of exfoliated flakes.

**Sample fabrication**. Samples were directly exfoliated onto quartz slides (20 mm×20 mm×0.5 mm), using traditional tape exfoliation method, for SHG and SPDC experiments on single flakes. For twisted stack samples, two flakes were initially exfoliated onto a quartz slide and a polydimethylsiloxane (PDMS) film, respectively. Subsequently, the selected flake on PDMS was transferred onto the other selected flake on quartz using an all-dry transfer method within a glovebox. Note that the exfoliated $NbOCl_2$ flakes, both on quartz and PDMS, exhibit highly anisotropic features facilitating the identification of crystallographic axes and control of twist angles.

**SHG experiment.** The SHG experiments were conducted in transmission geometry with the experimental setup illustrated in Supplementary Figure S1a. A tunable femtosecond laser (Coherent Chameleon Ultra, pulse width: 140 fs; repetition frequency: 80 MHz) was used as the pump source. A half-wave plate (HWP) before coupling into the fiber is used to control the pump power. Then the pump is coupled into a single-mode fiber (SF) and modulated to different linear polarizations with a polarization beam splitter (PBS), an 808-nm HWP. A long-pass (LP) filter (at 700 nm) is used to remove possible noise generated in fiber. The pump light is focused onto the sample by an objective (10×, N.A.=0.25) and the generated SHG signal is collected by another objective (10x, N.A.=0.25). The SHG signal is projected into different linear polarizations with a HWP and a PBS. After removing the pump light by a short-pass filter, the SHG photons are coupled into an SF and recorded by single photon detector (SPD, PerkinElmer, SPCM-AQRH-15-FC). Spectrum was measured using a spectrometer (Princeton Instruments, SP2500) cooled by liquid nitrogen. All the experiments were operated at room temperature.

**Biphoton correlation measurements.** The two-photon correlation properties were recorded with the setup illustrated in Supplementary Figure S3a. The pump light from a 404-nm continuous-wave laser (Toptica, DL Pro) is coupled into an SF, followed by modulation into different linear polarizations using a PBS and an HWP. The HWP before coupling into SF is used to control the pump power, while the band-pass (BP, Thorlabs, FBH405-10) after the first PBS is used to remove the possible noise generated in fiber. The pump is focused onto the sample with a 10x objective (N.A.=0.5). After removing the pump light with an LP filter (at 700 nm) and a BP filter (Thorlabs, FBH810-10), the signal and idler photons are projected into the same various linearly polarizations with an HWP and a PBS. Finally, signal and idler photons are coupled into a same multimode fiber (MF) and separated by the fiber beam splitter (FBS, splitting radio 50:50, Thorlabs, TG625R5F1A), followed by recording by two SPDs (PerkinElmer, SPCM-AQRH-15-FC) respectively. A Time-Correlated Single Photon Counting (TCSPC) module (qutools quTAG) is used to calculate the second-order correlation function and the coincidence of the biphotons.

**Polarization quantum state tomography.** The experimental setup for quantum state tomography is illustrated in Figure 4c. The pump source and focusing process are the same with that of above biphoton correlation experiments. While in this case, the signal and idler photons are separated by a polarization-independent beamsplitter (BS, splitting ratio 50:50) and individually projected into different polarization basis with a sequence of HWP, quarter-wave plate (QWP) and a PBS. This is followed by coupling the SPDC photons into MFs, detecting by SPDs and calculating the coincidence by TCSPC.

**Simulations of polarization-entangled states in twisted vdW stacks.** The quantum state of photon pairs generated at $z = z_0$ by SPDC could be represented as[28,47-50]:

$$|\psi(\theta, \alpha_j, z_0)\rangle = \sum_{lmn} \zeta_{lmn}(\theta, \alpha_j, z_0) \hat{a}_{i,m}^\dagger \hat{a}_{s,n}^\dagger e^{i\Delta k z_0}|0\rangle, \quad (1)$$

in which $\hat{a}_{s,n}^\dagger$ and $\hat{a}_{i,m}^\dagger$ are generation operators of signal ($s$) and idle ($i$) photons along $n$- and $m$-polarizations, respectively; $\zeta_{lmn}(\theta, \alpha_j, z_0) \propto \chi_{lmn}^{(2)}(\alpha_j, z_0) E_{p,l}(\theta, \alpha_j, z_0)$ is the nonlinear conversion coefficient which is proportional to the second-order nonlinear susceptibility of the material $\chi_{lmn}^{(2)}(\alpha_j, z_0)$ and the electric field amplitude of the pump laser along $l$-polarization $E_{p,l}(\theta, \alpha_j, z_0)$; $\Delta k(\alpha_j, z_0) = k_p(\alpha_j, z_0) - k_s(\alpha_j, z_0) - k_i(\alpha_j, z_0)$ is the wave vector mismatch. $\theta$ is the polarization angle of the pump laser; $\alpha_j$ is the twist angle of the corresponding $j$-th layer around $z$ axis. The overall quantum state of photon pairs generated by multilayer twisted NbOCl$_2$ should be derived by integrating Eq. (1) within the thin-films region:

$$|\Psi(d_1, d_2, \cdots, d_m, \alpha_1, \alpha_2, \cdots, \alpha_m, \theta)\rangle = \int_0^{d_1+d_2+\cdots+d_m} \frac{\partial |\psi(\theta, \alpha_j, z)\rangle}{\partial z} dz. \quad (2)$$

where $d_j$ ($j = 1, 2, \cdots, m$) represents the thicknesses of each stacked layer.

The refractive indices and the second-order nonlinear susceptibility matrix of the $j$-th layer rotate along with the twist angle:

$$\begin{bmatrix} n_{zz,v} & n_{zx,v} & n_{zy,v} \\ n_{xz,v} & n_{xx,v} & n_{xy,v} \\ n_{yz,v} & n_{yx,v} & n_{yy,v} \end{bmatrix} = \mathbf{R}_n^T(\alpha_j) \begin{bmatrix} n_{a,v} & 0 & 0 \\ 0 & n_{b,v} & 0 \\ 0 & 0 & n_{c,v} \end{bmatrix} \mathbf{R}_n(\alpha_j), \quad (3)$$

$$\boldsymbol{\chi}_j^{(2)}(\alpha_j) = \mathbf{R}_{\chi,\text{ii}}^T(\alpha_j) \boldsymbol{\chi}^{(2)} \mathbf{R}_{\chi,\text{i}}(\alpha_j), \quad (4)$$

where the subscripts $x$ corresponds to $H$ axis and $y$ corresponds to $V$ axis; $a, b$ and $c$ are NbOCl$_2$ crystal axes; $v = p, s, i$ corresponds to pump laser, signal and idle photons, respectively; the superscript $T$ represents the transpose; $\mathbf{R}_n$ is the rotation matrix of the refractive indices, and $\mathbf{R}_{\chi,\text{i}}$, $\mathbf{R}_{\chi,\text{ii}}$ are rotation matrices of the nonlinear susceptibility:

$$\mathbf{R}_n(\alpha_j) = \mathbf{R}_{\chi,\text{ii}}(\alpha_j) = \begin{bmatrix} 1 & 0 & 0 \\ 0 & \cos\alpha_j & \sin\alpha_j \\ 0 & -\sin\alpha_j & \cos\alpha_j \end{bmatrix}, \quad (5)$$

$$\mathbf{R}_{\chi,\text{i}}(\alpha_j) = \begin{bmatrix} 1 & 0 & 0 & 0 & 0 & 0 \\ 0 & \cos^2\alpha_j & \sin^2\alpha_j & \sin\alpha_j\cos\alpha_j & 0 & 0 \\ 0 & \sin^2\alpha_j & \cos^2\alpha_j & -\sin\alpha_j\cos\alpha_j & 0 & 0 \\ 0 & -\sin 2\alpha_j & \sin 2\alpha_j & \cos 2\alpha_j & 0 & 0 \\ 0 & 0 & 0 & 0 & \cos\alpha_j & -\sin\alpha_j \\ 0 & 0 & 0 & 0 & \sin\alpha_j & \cos\alpha_j \end{bmatrix}. \quad (6)$$

Specifically, for orthogonally stacked bilayers, we set $\alpha_1 = 0°$ and $\alpha_2 = 90°$; for arbitrary-angle twisted bilayers (Supplementary Figure S11), $\alpha_1$ is set to $0°$ and $\alpha_2$ is swept from $0°$ to $180°$; for tri-layer stacked thin films configuration (Supplementary Figures S12 and S13), we fixed $\alpha_1 =$

0° and $\alpha_3 = 90°$, then also twisted the second layer $\alpha_2$ from 0° to 180°.

The generated entangled states in Eq. (2) can be written using four orthogonal polarization bases:

$$|\Psi\rangle = \frac{a_1 e^{i\phi_1}|HH\rangle + a_2 e^{i\phi_2}|VV\rangle + a_3 e^{i\phi_3}|HV\rangle + a_4 e^{i\phi_4}|VH\rangle}{\sqrt{|a_1|^2 + |a_2|^2 + |a_3|^2 + |a_4|^2}} \quad (7)$$

The density matrix of the state is $\boldsymbol{\rho} = |\Psi\rangle\langle\Psi|$. Then, by introducing a non-Hermitian matrix $\mathbf{R} = \boldsymbol{\rho}\boldsymbol{\sigma}\boldsymbol{\rho}^T\boldsymbol{\sigma}$, where

$$\boldsymbol{\sigma} = \begin{pmatrix} 0 & 0 & 0 & -1 \\ 0 & 0 & 1 & 0 \\ 0 & 1 & 0 & 0 \\ -1 & 0 & 0 & 0 \end{pmatrix}, \quad (8)$$

the concurrence of the state can be determined by[51]:

$$C = \max\{0, \sqrt{r_1} - \sqrt{r_2} - \sqrt{r_3} - \sqrt{r_4}\}, \quad (9)$$

in which $r_1 \geq r_2 \geq r_3 \geq r_4$ are eigenvalues of $\mathbf{R}$ in decreasing order. The fidelity of the generated entangled state can be derived by[52]:

$$F^\pm = \text{Tr}(\boldsymbol{\rho}|\boldsymbol{\Phi}^\pm\rangle\langle\boldsymbol{\Phi}^\pm|), \quad (10)$$

where $|\boldsymbol{\Phi}^\pm\rangle = 1/\sqrt{2}\,(|HH\rangle \pm |VV\rangle)$ are the Bell states.

For the qutrit characterization of tri-layer stacks, we replaced the state definition Eq. (7) as:

$$|\Psi_t\rangle = \frac{a_1 e^{i\phi_1}|HH\rangle + a_2 e^{i\phi_2}|VV\rangle + a_3 e^{i\phi_3}|\xi\rangle}{\sqrt{|a_1|^2 + |a_2|^2 + |a_3|^2}}, \quad (11)$$

in which $|\xi\rangle = 1/\sqrt{2}\,(|HV\rangle + |VH\rangle)$. And in this case, $|\boldsymbol{\Phi}^\pm\rangle$ is replaced by the qutrit state $|\boldsymbol{\Phi}_t^+\rangle = 1/\sqrt{3}\,(|HH\rangle + |VV\rangle + |\xi\rangle)$ in Eq. (10) to determine the fidelity $|F_t^+\rangle$ of the generated qutrit.

**Data availability**

All relevant data are available in the main text, in the Supporting Information, or from the authors upon reasonable request.

**Acknowledgement**

This work is financially supported by the National Research Foundation, Prime Minister's Office, Singapore under Competitive Research Program Award NRF-CRP26-2021-0004. Q.Z. acknowledges the financial support from the Research Funds of Hangzhou Institute for Advanced Study, UCAS (A05006C019014). X.F.R. acknowledges the financial support from the National Key Research and Development Program of China (2022YFA1204704), the Innovation Program for Quantum Science and Technology (2021ZD0303200, 2021ZD0301500), the National Natural Science Foundation of China (NSFC) (62061160487, T2325022, U23A2074, 62205325), the CAS Project for Young Scientists in Basic Research (No. YSBR-049), Key Research and Development Program of Anhui Province (2022b1302007) and the Fundamental Research Funds for the Central Universities. This work was partially carried out at the USTC Center for Micro and Nanoscale Research and Fabrication. Q.Z. acknowledges the financial support from the Research Funds of Hangzhou Institute for Advanced Study, UCAS (A05006C019014, B03006C01600407) and Zhejiang Provincial Natural Science Foundation (Grant No. LQ24H180008).

**Author contributions**
Q.G. conceived and coordinated this work. Q.G. and Q.Z. synthesized the crystals, conducted structural and optical characterizations. Q.G. fabricated all the samples for experiments. Y.K.W. and X.F.R. conducted the SHG and SPDC experiments. D.Z. and A.A. did the SPDC theoretical modeling on stacked samples. Q.G. analyzed the data and drafted the manuscript. All authors discussed the results and contributed to the manuscript. Q.G. and C.W.Q. supervised the project.

**Competing interests**
The authors declare no interest of conflict.